\documentclass[aps,prd,superscriptaddress,preprint,12pt,nofootinbib]{revtex4} 
\usepackage{amsmath,amsfonts,amssymb,graphicx,natbib,bm}
\usepackage{color}
\usepackage{slashed}    
\usepackage{hyperref}
\usepackage{breakurl}
\usepackage{color}
\usepackage{url}
\usepackage{cancel}
\usepackage{slashed}
\usepackage{multirow}
\usepackage{rotating}

\newcommand{\be}{\begin{eqnarray*}}
\newcommand{\ee}{\end{eqnarray*}}

\newcommand{\bee}{\begin{eqnarray}}
\newcommand{\eee}{\end{eqnarray}}
\newcommand{\beeq}{\begin{equation}}
\newcommand{\eeq}{\end{equation}}

\usepackage{xspace}

\begin{document}

\preprint{IPPP/16/78, DCPT/16/156, LAPTH-036/16, PITT-PACC-1606}

\title{Invisible Decays in Higgs Pair Production}

\author{Shankha Banerjee}
\affiliation{LAPTH, Univ. de Savoie, CNRS, B.P.110, F-74941 Annecy-le-Vieux, France}

\author{Brian Batell}
\affiliation{Pittsburgh Particle Physics, Astrophysics, and Cosmology Center,
Department of Physics and Astronomy, University of Pittsburgh, Pittsburgh, USA}

\author{Michael Spannowsky}
\affiliation{Institute for Particle Physics Phenomenology, Department of Physics, Durham University, DH1 3LE, UK}

\begin{abstract}
Observation of Higgs pair production is an important long term objective of the LHC physics program as it will shed light on the scalar potential of the Higgs field and the nature of electroweak symmetry breaking. While numerous studies have examined the impact of new physics on di-Higgs production, little attention has been given to the well-motivated possibility of exotic Higgs decays in this channel. Here we investigate the consequences of exotic invisible Higgs decays in di-Higgs production. We outline a search sensitive to such invisible decays in the $b\bar b+{\not \!\! E}_T$ channel. We demonstrate that probing invisible branching ratios of order 10$\%$ during the LHC's high-luminosity run will be challenging, but in resonance enhanced di-Higgs production, this final state can become crucial to establish the existence of physics beyond the Standard Model at collider energies. We also briefly discuss the outlook for other exotic Higgs decay modes and the potential to observe such exotic decays in the di-Higgs channel. 
\end{abstract}

\maketitle
\newpage

\baselineskip=15pt

\baselineskip=18pt

\small

\newpage

\section{Introduction}

The di-Higgs channel has emerged as a holy grail at high energy colliders such as the large hadron collider (LHC). This channel provides a means to directly probe the Higgs cubic coupling, one of the last unmeasured parameters of the Standard Model (SM), and more broadly explore the shape of the scalar potential and in turn the dynamics underlying electroweak symmetry breaking.  
However, achieving discovery in this channel and ultimately obtaining a precise measurement of the Higgs self coupling is known to be challenging at hadron colliders. 
This is primarily due to the small di-Higgs production rate, which is a consequence of the partial cancellation between the ``box'' and ``triangle'' diagrams~\cite{Glover:1987nx,Eboli:1987dy,Plehn:1996wb,Dawson:1998py},\cite{Li:2013rra,Dicus:2015yva,Dawson:2015oha}.
Nevertheless, it is expected that the LHC and a future 100 TeV hadron collider will be able to perform a measurement of the self coupling at the 10-50$\%$ level depending on the size of the dataset, proton energy, and ability to control background systematics~\cite{Baur:2002rb,Baur:2002qd,Baur:2003gp,Dolan:2012rv,Papaefstathiou:2012qe,Baglio:2012np,Goertz:2013kp,Barr:2013tda,Dolan:2013rja,deLima:2014dta,Englert:2014uqa,Liu:2014rva,Barr:2014sga,Azatov:2015oxa,Papaefstathiou:2015iba,Dolan:2015zja,Behr:2015oqq,Kling:2016lay,Contino:2016spe}. 

Given the importance of the di-Higgs channel along with the numerous theoretical and empirical motivations to go beyond the Standard Model (BSM), it becomes necessary to understand how new physics can manifest in this channel and optimize the search strategy appropriately. An important example in this regard is the case of a di-Higgs resonance: a new heavy particle produced in the $s$ channel that decays to a pair of Higgs bosons. Besides enhancing the di-Higgs production cross section, the resonance will alter the kinematics of the $hh$ system and the Higgs decay products, thus warranting a different experimental approach in comparison to the one employed for the case of the SM. We will show that if such a scenario is realised in nature, e.g. in form of a Higgs portal to a dark sector, searches in di-Higgs final state can be the first process to evidence invisible Higgs decays at the LHC, thus breaking with the paradigm that vector-boson-fusion (VBF) induced Higgs production yields the best limits on the branching ratio $h \to \mathrm{invisible}$ \cite{Eboli:2000ze,Bernaciak:2014pna}.

Studies of new physics effects on the di-Higgs channel have to date focused on modifications to the {\it production} of Higgs boson pairs. Examples include the di-Higgs resonance mentioned above, as well as non-resonant contributions due to loops of new light colored particles or higher-dimensional operators~~\cite{Plehn:1996wb,Djouadi:1999rca,Belyaev:1999mx,BarrientosBendezu:2001di,Pierce:2006dh,Arhrib:2009hc,Asakawa:2010xj,Kribs:2012kz,Dawson:2012mk,Dolan:2012ac,Goertz:2013kp,Cao:2013si,Han:2013sga,Nishiwaki:2013cma,Haba:2013xla,Enkhbat:2013oba,Kumar:2014bca,Chen:2014xra,Chen:2014xwa,Chen:2014ask,Cao:2014kya,Azatov:2015oxa,Grober:2015cwa,Wu:2015nba,Enkhbat:2015bca,Lu:2015jza,Dawson:2015oha,Etesami:2015caa,He:2015spf,Carvalho:2015ttv,Batell:2015zla,Lu:2015qqa,Batell:2015koa,Dawson:2015haa,Cao:2015oaa,Huang:2015tdv,Agostini:2016vze,Grober:2016wmf,Kang:2016wqi}. However, it has long been recognized that new physics can easily affect the {\it decays} of the Higgs, and indeed the subject of exotic Higgs decays is under active theoretical and experimental investigation~\cite{Curtin:2013fra}. Because the single Higgs production rate is much larger than the di-Higgs production rate, one generically expects exotic Higgs decays to first manifest in channels involving production of a single Higgs particle, unless perhaps the di-Higgs production rate is significantly enhanced by one of the mechanisms mentioned above. 
Nevertheless, it is important to consider the potential implications of exotic decays in the di-Higgs channel for several reasons.  
First, the current branching ratio limits for a variety of exotic Higgs decay channels are very weak, and can easily be in the 
10-50$\%$ range in some cases. Second, the di-Higgs final state will be altered if one or both of the Higgs decays into the exotic channel, thus leading to a different experimental signature and necessitating a different search strategy. 

With this motivation, in this paper we investigate the implications of the invisible decay mode, 
$h \rightarrow \, {\not\!\! E}_T$, on the di-Higgs channel. While there are many possibilities for exotic Higgs decay modes, the invisible channel is particularly well-motivated on several counts. 
As we will review below, there are numerous BSM scenarios that predict an invisibly decaying Higgs. For example, the Higgs may provide a portal to the dark sector and thus decay into light dark matter particles. 
Another reason for considering the invisible channel is simply that it is a generic possibility; even if the new light particles resulting from the exotic Higgs decay are not absolutely stable, they may be long-lived on collider scales, or cascade decay to other long lived particles, resulting in a missing energy signature. We outline a search in the $b\bar b+ {\not \!\!E_T}$ channel and examine the LHC prospects both for the case of SM Higgs pair production as well as for enhanced production through a di-Higgs resonance. 
Such a di-Higgs resonance  is well-motivated in its own right from several perspectives. As one example, Higgs portal models might allow for a strong first-order phase transition, a necessary ingredient to a viable baryogenesis mechanism \cite{Assamagan:2016azc}. The simplest Higgs portal model in which the SM is extended by a real singlet scalar, the so-called xSM \cite{Profumo:2007wc,Barger:2007im,Profumo:2014opa,Kotwal:2016tex}, predicts a scalar resonance of ${\mathcal O}(500)$ GeV to enable a first-order phase transition, which can manifest at colliders as in the di-Higgs channel. 
Although probing the SM di-Higgs production rate will be challenging at the high-luminosity LHC, our search has the potential to probe phenomenologically viable invisible branching ratios, particularly if the di-Higgs cross section is moderately enhanced by new physics or if a di-Higgs resonance is present. 
 
It is worth noting that the signature we propose is essentially the ``mono-Higgs'' signature that has been suggested as a probe of certain dark matter scenarios in Refs.~\cite{Carpenter:2013xra,Berlin:2014cfa,Basso:2015aee}, although these studies have little overlap with our investigation here. In particular, certain cuts employed in the $b\bar b+ {\not \!\!E_T}$ search in Ref.~\cite{Carpenter:2013xra} are not optimized to di-Higgs production, notably the hard ${\not \!\!E_T}$ cut. 
Furthermore, Ref.~\cite{Berlin:2014cfa} does not consider Higgs pair production, even as a potential background, as they are interested in dark matter models with an enhanced mono-Higgs signal. As we will discuss in detail below, the existing LHC searches in mono-Higgs channels do not provide constraints on our scenario that are competitive with direct searches for di-Higgs production in standard channels. 

In the next section we provide a brief review of the theoretical motivation and experimental status of invisible Higgs decays. In Section~\ref{sec:SM} we discuss the case of standard di-Higgs production, describing our search strategy in the $b\bar b+ {\not \!\!E_T}$ channel and its prospects for the LHC, while in Section~\ref{sec:resonance} we consider the case of a di-Higgs resonance. In Section~\ref{sec:exotic} we provide some preliminary discussion on other exotic decay channels in Higgs pair production. Our conclusions and outlook are presented in Section~\ref{sec:conclusion}.

\section{Invisible Higgs Decays}
\label{sec:invisible}

The Higgs boson of the SM is very narrow, having a width of about 4 MeV. This makes it highly susceptible to new exotic decay modes. Indeed, if any new light particles couple to the Higgs with a strength that is comparable to the bottom quark Yukawa coupling, $\sim 1/60$, the Higgs can have a sizable exotic branching ratio to these new light states. 

The subject of invisible Higgs decays has a long history and dates back to Ref.~\cite{Shrock:1982kd}. There are a number of reasons to consider the invisible decay as a primary exotic decay mode. First, even without any particular new physics motivation, invisible decays appear to be a generic possibility if the Higgs couples to new light states. Provided the new light particles are weakly interacting with matter, stable, metastable with a macroscopic decay length that exceeds the detector size ${\cal O}(10\, {\rm m})$, or cascade decay to other (meta)stable states, the signature of the Higgs at high energy colliders will involve missing transverse energy. More importantly, there are a number of new physics motivations for considering exotic invisible Higgs decays. The Higgs can mediate interactions between dark matter and the SM, and allow its decay to dark matter~\cite{Silveira:1985rk,McDonald:1993ex,Burgess:2000yq}, or to more general hidden/dark sectors~\cite{Binoth:1996au,Schabinger:2005ei,Strassler:2006im,Patt:2006fw,Antusch:2015mia}. The Higgs could decay to light sterile neutrinos that are long lived~\cite{Graesser:2007pc,Graesser:2007yj}, or pseudo Nambu-Goldstone bosons such as axions or Majorons~\cite{Shrock:1982kd,Diaz:1998zg}. The Higgs may also decay to the Lightest Supersymmetric Particle (LSP) in supersymmetric extensions of the SM~\cite{Gunion:1988yc}, or to Kaluza-Klein states in extra-dimensional theories~\cite{Beauchemin:2004zi,Diener:2013xpa}. These are but a small sample of motivated new physics scenarios predicting an invisible Higgs, but there are many others, for which we refer the reader to the review article~\cite{Curtin:2013fra}.

The invisible Higgs branching ratio, 
\begin{equation}
{\rm Br}_{\rm inv} \equiv {\rm Br}(h\rightarrow {\not\!\! E}_T),
\label{eq:Br-inv}
\end{equation}
can be constrained at the LHC in two ways~\cite{Khachatryan:2014jba}. The first is from global fits to the Higgs couplings, which rely on some assumptions about the new physics modifications to Higgs properties. The second is direct searches for the $h\rightarrow {\not\!\! E}_T$ signature~\cite{Choudhury:1993hv,Frederiksen:1994me,Eboli:2000ze,Martin:1999qf,Godbole:2003it,Davoudiasl:2004aj,Bai:2011wz,Djouadi:2012zc,Bernaciak:2014pna,Antusch:2015gjw}, which can be searched for in the mono-jet ($hj$), VBF($hjj$), and associated $Vh$ channels~\cite{Aad:2015txa,Aad:2014iia,Aad:2015uga,Chatrchyan:2014tja}. The combined constraints on the invisible branching ratio are currently in the range ${\rm Br}_{\rm inv}\lesssim 25\%-50\%$, depending on the assumptions about the other Higgs couplings~\cite{Aad:2015pla,Khachatryan:2014jba}. 
In the future, the LHC will be able to probe ${\rm Br}_{\rm inv} $ down to the order 5$\%$ level with 3000 fb$^{-1}$ at the high luminosity run. 

In the next section we will examine the potential for the LHC to observe invisible Higgs decays in Higgs pair production. However, it is already clear that, unless the di-Higgs production rate is significantly enhanced by new physics, it is very likely that an exotic invisible decay of the Higgs will first be discovered in single Higgs production channels. This does not however diminish the importance of studying the invisible decays in the di-Higgs channel. If invisible Higgs decays are indeed realized in nature, then the di-Higgs channel will offer an important additional channel to confirm and study their properties. Furthermore, as argued in the introduction, one of the primary interests to understand the di-Higgs channel is to probe the scalar potential, and if ${\rm Br}_{\rm inv}$ is sizable it will be necessary to study invisible Higgs decays in this channel to this end.

\section{Standard Di-Higgs Production}
\label{sec:SM}

We now investigate in detail the prospects at the LHC ($\sqrt{s} = 14$ TeV) to observe exotic invisible Higgs boson decays in the di-Higgs production channel.  
In this section we consider only SM di-Higgs production modes, which proceed at the partonic level via the dominant box and triangle loop contributions to $gg\rightarrow hh$~\cite{Glover:1987nx,Eboli:1987dy,Plehn:1996wb,Dawson:1998py}. Since the Higgs discovery significant advances have been made in higher order computations of di-Higgs production~\cite{Baglio:2012np,Shao:2013bz,Grigo:2013rya,Chen:2013emb,deFlorian:2013jea,Maltoni:2014eza,Frederix:2014hta,Grigo:2014jma,deFlorian:2015moa,Grigo:2015dia,Cao:2015oxx,deFlorian:2016uhr,Bian:2016awe}. 
The total cross section has now been computed at Next-to-leading order (NLO) including the full top quark mass dependence~\cite{Borowka:2016ehy,Borowka:2016ypz} and differentially at Next-to-
next-to-leading order (NNLO) in the heavy top mass limit~\cite{deFlorian:2016uhr}. In our study, we will use the NNLO prediction from Ref.~\cite{deFlorian:2016uhr} for $\sqrt{s} = 14$ TeV:
\begin{equation}
\sigma^{hh}_{\rm NNLO} = 37.52(4)^{+5.2\%}_{-7.6\%} ~{\rm fb}.
\label{eq:SM-XS}
\end{equation}

We will be interested in di-Higgs events in which one of the Higgses decays invisibly, with a new physics invisible branching ratio,~Eq.~(\ref{eq:Br-inv}),
while the other Higgs decays to visible SM particles. The Higgs boson decays to several clean channels, notably to two photons and four leptons, which suggest searching for the $2\gamma+{\not\!\! E}_T$, $4\ell+{\not\!\! E}_T$ final states as a sign of di-Higgs production. However, the small branching ratios of these decays coupled with the small SM di-Higgs production cross section of Eq.~(\ref{eq:SM-XS}) makes it challenging to exploit these clean channels due to the resulting low signal rate. We will therefore not consider these clean channels here, although they could be quite interesting if the di-Higgs production rate is significantly enhanced by new physics (see Sec.~\ref{sec:resonance} for further discussion). 
Another potentially interesting channel for di-Higgs is $WW^*+{\not\!\! E}_T$, due to the sizable $h \rightarrow WW^*$ branching ratio. However, leptonic $W$ decays make reconstruction of the Higgs challenging, particularly due to the additional missing energy in the event coming from the other invisible Higgs, while the fully hadronic channel must contend with a large SM background. Similar considerations apply to the $2\tau+{\not\!\! E}_T$ channel.

Given these considerations, our focus in this paper will therefore be on the $b\bar b+{\not\!\! E}_T$ channel, i.e.,
\begin{equation}
p p \rightarrow  h h + X \rightarrow (b  \bar b) ({\not\!\! E}_T) + X.
\label{eq:bbMET}
\end{equation}
The obvious advantage of this channel is that it provides the largest possible signal rate, allowing us to impose hard cuts to separate the di-Higgs signal from the various backgrounds. 
While this channel will be the primary focus of our study here, we stress that it will be important to utilize the other channels mentioned above. Exploiting the other channels may significantly improve the sensitivity, particularly in the case of enhanced di-Higgs production, and will also allow for a more robust interpretation of the exotic invisible Higgs decay. We leave this important work to future study.

We now turn to our analysis of the $b\bar b+{\not \!\! E}_T$ channel. Signal and background event samples are generated with \texttt{MadGraph5\_aMC@NLO}~\cite{Alwall:2014hca}. The tree level samples are generated with the Parton distribution functions (PDF) set \texttt{CTEQ6L1}~\cite{Pumplin:2002vw} and passed through \texttt{Pythia 6}~\cite{Sjostrand:2006za} for showering and hadronization. For the loop level samples, generated with the \texttt{NN23LO1} PDF set~\cite{Ball:2013hta,Ball:2014uwa}, we first decay the particles using \texttt{MadSpin}~\cite{Frixione:2007zp,Artoisenet:2012st} and then shower and hadronize the samples in the \texttt{Pythia 8} framework. We use \texttt{Delphes 3}~\cite{deFavereau:2013fsa}~\footnote{We thank Shilpi Jain, Alexandre Mertens and Michele Selvaggi for help in understanding intricacies about the $b$-jet tagging and treating a new particle as missing energy in the Delphes framework.} with the default ATLAS card for fast detector simulation. The jets are formed using the anti-$k_T$ algorithm~\cite{Cacciari:2008gp} in the \texttt{FASTJET}~\cite{Cacciari:2011ma} framework with the parameter $R = 0.5$.
We have also performed a cross-check of the signal events using \texttt{Herwig 7}~\cite{Bellm:2015jjp,Bahr:2008pv} finding good agreement at all stages in the analysis chain.

The presence of the invisible branching ratio Eq.~(\ref{eq:Br-inv}) has two effects on the production rate of the signal and the background channels involving a single Higgs. First, the signal and some of the background channels contain a Higgs that decays invisibly. These channels require the inclusion of a scale factor $\sim {\rm Br}_{\rm inv}$ in the rate estimate. 
Second, the new invisible decay mode has the effect of diluting the branching ratios of the SM Higgs decay modes. Thus, the rates of the signal and those backgrounds which contain the decay $h\rightarrow b\bar b$ must be scaled by a dilution factor $\sim(1-{\rm Br}_{\rm inv})$. The signal rate should of course be scaled by both factors. 
In addition, we will include in this section a signal strength factor for di-Higgs production, 
\begin{equation}
\mu_{hh} \equiv \frac{\sigma^{hh}}{\sigma_{\rm SM}^{hh}}.
\end{equation}
We will thus assess the sensitivity of our search strategy in the model-independent ${\rm Br}_{\rm inv} - \mu_{hh}$ parameter space. 

It is worth emphasizing here that the standard di-Higgs signatures, such as $2b2\gamma$, $2b2W$, $2b2\tau$, $4b$, and so forth, will become more challenging to discover if a non-zero invisible branching ratio, or any other exotic branching ratio for that matter, is realized in nature. Since the branching ratios in the standard channels are diluted by the factor $\sim(1-{\rm Br}_{\rm inv})$, the overall signal rate in these channels will be smaller by a factor $\sim(1-{\rm Br}_{\rm inv})^2$ compared to the SM. 

There are several important backgrounds to the $b\bar b+{\not \!\! E}_T$ di-Higgs signal that we will need to consider. We distinguish these based on their scaling with ${\rm Br}_{\rm inv}$:
\begin{itemize}
\item Scale $\propto {\rm Br}_{\rm inv}$: This class involves production of a single Higgs particle that subsequently decays invisibly, notably $Zh$ production in which $Z\rightarrow 2b $ and $h\rightarrow {\not \!\! E}_T$.
\item Scale $\propto  (1-{\rm Br}_{\rm inv})$: This class similarly involves production of a single Higgs particle, which instead decays to bottom quarks. The specific examples include associated production, $Wh$ followed by $W\rightarrow \ell \nu$, $h\rightarrow 2b$ as well as $Zh$ followed by $Z\rightarrow 2\nu/2\ell$, $h \rightarrow 2b$. Large ${\not \!\! E}_T$ results when the leptons are missed in the detector. 
\item Independent of ${\rm Br}_{\rm inv}$: This class of backgrounds includes all processes that do not involve production of a Higgs particle. The prominent examples in this group are $Z b\bar b$ (no $h$), and
$W b\bar b$ (no $h$), where the $Z$ or $W$ decays to neutrinos and/or leptons, as well as leptonic or semi-leptonic $t\bar t$, where again the leptons are not identified. 
\end{itemize}
For the associated $Zh$ production we include both the Born-level process as well as the one-loop process $gg\rightarrow Zh$. We generate both these processes at LO and then scale the numbers by the respective $K$-factors, see Table~\ref{tab:cutflow}. In addition, we also take into account a correction factor for Br($h \to b\bar{b}$) in order to mimic the inclusion of higher order corrections. The same tactic is followed for the $Wh$ process. Furthermore, the cross sections for the $t\bar t$, $Wb\bar b$ and $Zb\bar b$ channels are very large, ranging from $10^2 -10^3$ pb, leading us to impose hard selections at the generator level to generate sufficient statistics in the high $p_T$ regime. All generator level cuts are, however, still looser than those employed in our search, to be described below. For all of the backgrounds we apply uniform $K$-factors to the production cross sections. The references for these $K$-factors for the individual backgrounds are listed in Table~\ref{tab:cutflow}. 

In addition, there are several potential backgrounds that are not feasible to simulate in practice, and which we have therefore not included in our study. An example is QCD production of $b \bar b$ with large ``fake'' missing transverse energy arising from imperfect detector resolution and missed visible objects. The cross section before cuts for this process is enormous, $\cal{O}(\mu {\rm b})$, and thus a data driven approach for this background will be required. Despite its large rate, we expect that the hard $h\rightarrow {\not \!\! E}_T$ cut required in our search (see below) should be sufficient to eliminate this background \cite{Aaboud:2016nwl,Khachatryan:2015wza}. Similarly, potential backgrounds from $Vjj$, $Vbj$, with $V = W,Z$ can potentially mimic the signal if the jet fakes a $b$ and any leptons from the $V$ decay are not identified. However, despite the large cross sections, the small ($j\rightarrow b$) fake rate, typically of order $10^{-2}$, suggests that these backgrounds will be subdominant to the $Vb\bar b$ processes that we simulate in this study. For the sake of completeness, we also consider the single top background, finding that this background is sub-dominant and hence we do not include it in the table.

\begin{figure*}
\begin{center}
\includegraphics[width=0.45\columnwidth]{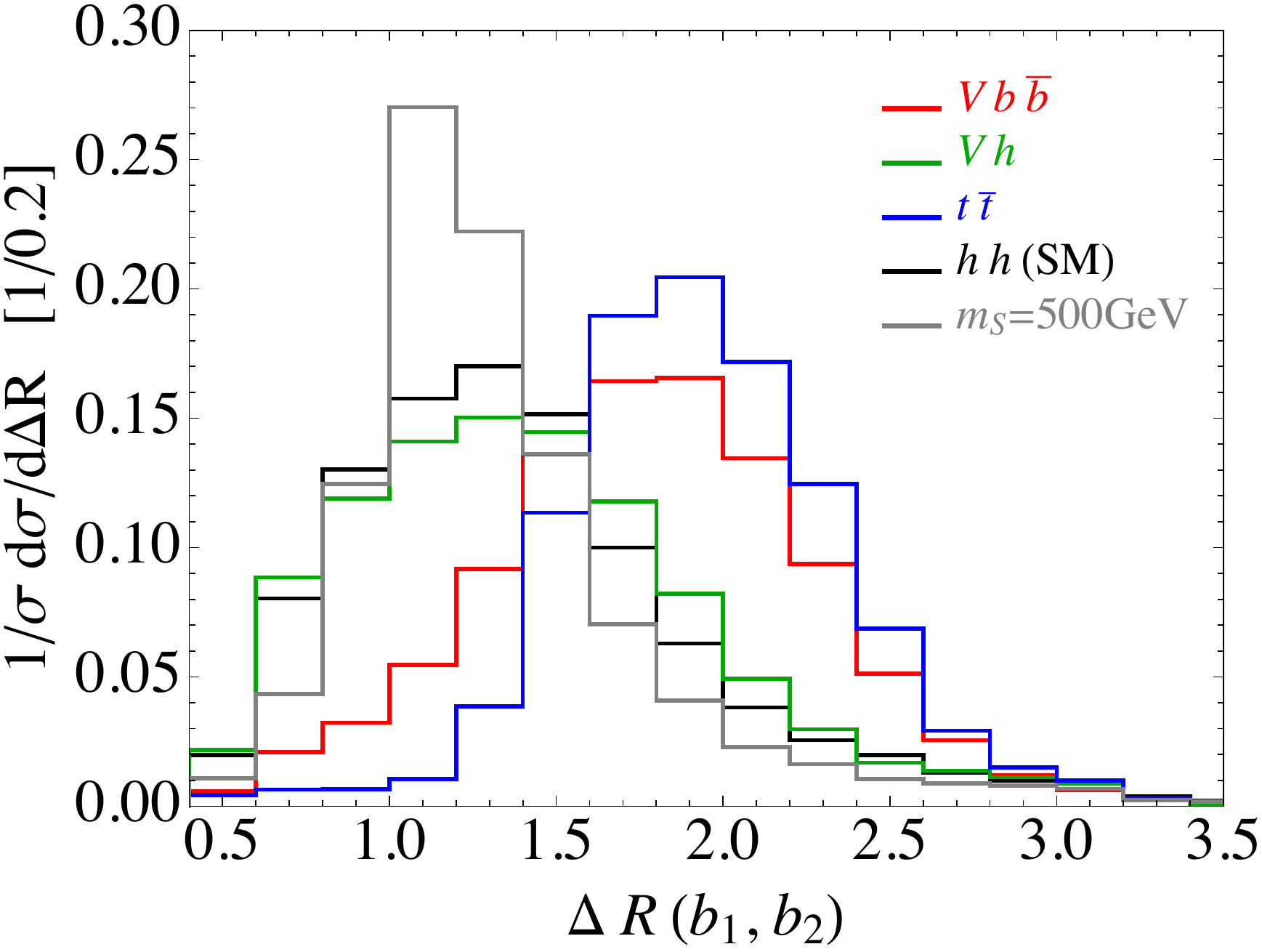}~~~~
\includegraphics[width=0.45\columnwidth]{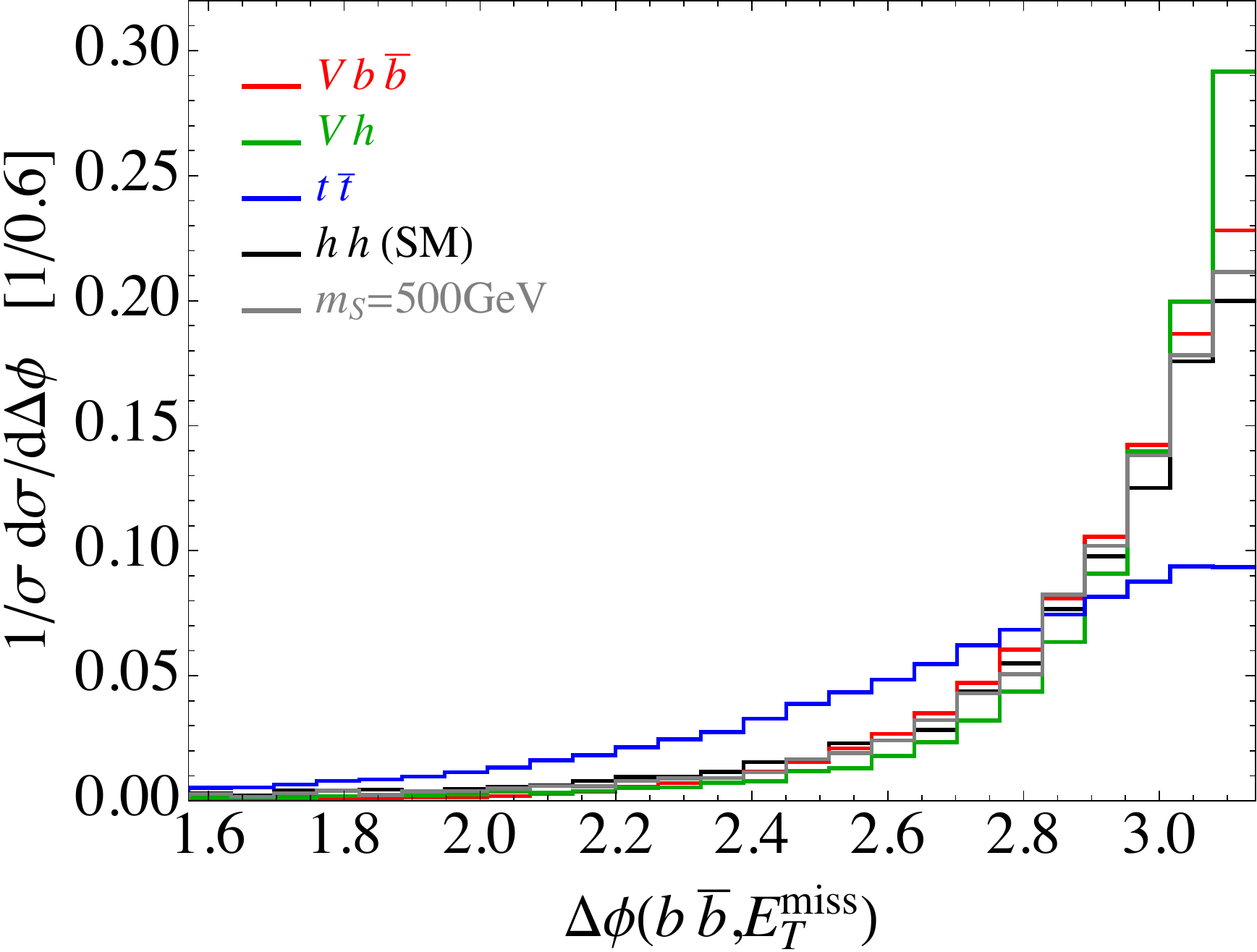}\\
~\\
\includegraphics[width=0.47\columnwidth]{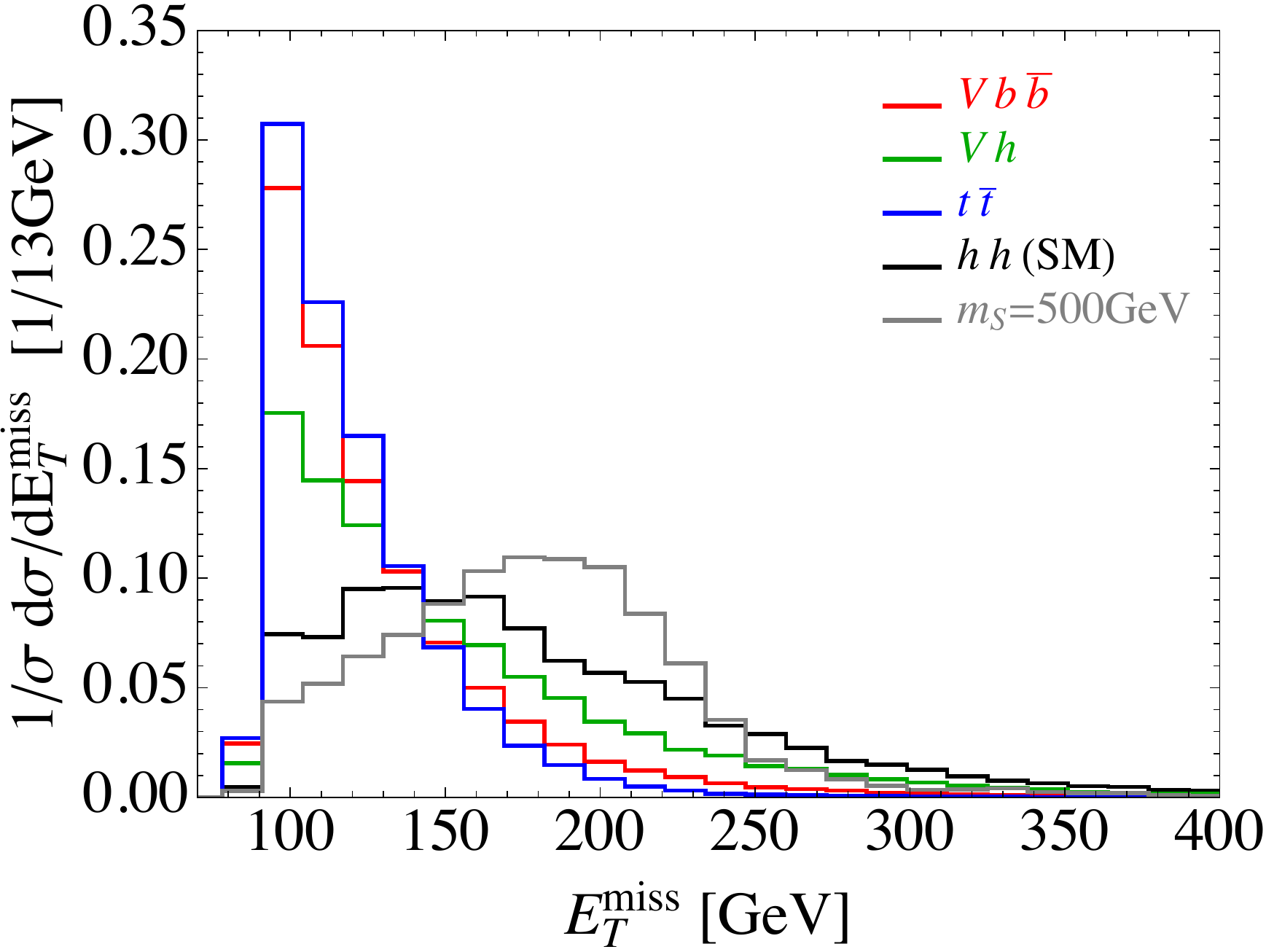}\\
\caption{Kinematic distributions for the variables $\Delta R(b_1,b_2)$, $\Delta \phi(b \bar b, {\not \!\! E}_T)$, and ${\not \!\! E}_T$ after the first selection of two $b$-jets. 
Here we have fixed ${\rm Br}_{\rm inv}=0.2$.
}
\label{fig:kinematics1}
\end{center}
\end{figure*}

We now turn to our search in the $b\bar b+{\not \!\! E}_T$ channel. Taking cue from a recent CMS search~\cite{CMS-PAS-EXO-16-012} for mono Higgs, we employ a trigger cut on $\slashed{E}_T$, \textit{viz.} $\slashed{E}_T > 90$ GeV. We then proceed by selecting two $b$-jets with $p_T > 35$ GeV and $|\eta| < 2.5$. 
We allow at most one additional jet with $p_T > 35$ GeV. Furthermore, events containing leptons with $p_T > 10$ GeV and $|\eta| < 2.5$ are vetoed. 
In Figure~\ref{fig:kinematics1} we show several kinematic distributions of the signal and the backgrounds following this first basic selection. To reconstruct the visible Higgs, the invariant mass of the two hardest $b$-jets is required to lie in a window around the Higgs mass in the range $115\,{\rm GeV} < m_{bb} < 135\,{\rm  GeV}$. 
To separate the di-Higgs signal from the dominant $t\bar t$, $Z b \bar b$, $Wb \bar b$ backgrounds, we next apply a cut on the angular separation of the two $b$-jets, demanding $0.4 < \Delta R(b_1,b_2)< 2$.
Further discrimination of the signal and the $t\bar t$ background is achieved by demanding the $b \bar b$ system and the missing transverse momentum to be separated in the azimuthal direction, $\Delta \phi(b \bar b, {\not \!\! E}_T) > 2.5$. In the final step of our initial selections, we place a hard cut on the missing transverse energy in the event, requiring ${\not \!\! E}_T > 160$ GeV. 

\begin{figure*}
\begin{center}
\includegraphics[width=0.45\columnwidth]{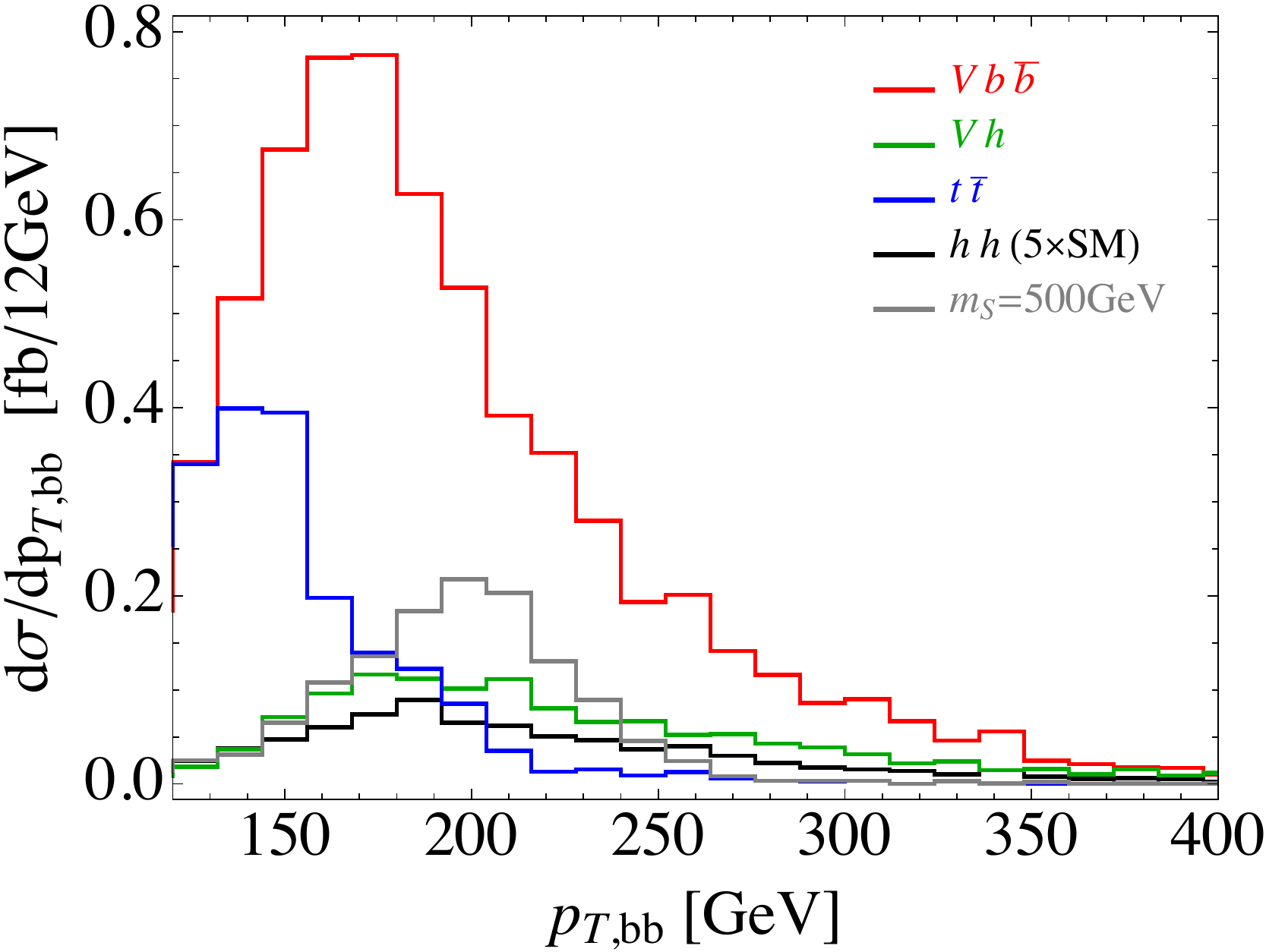}~~~~
\includegraphics[width=0.45\columnwidth]{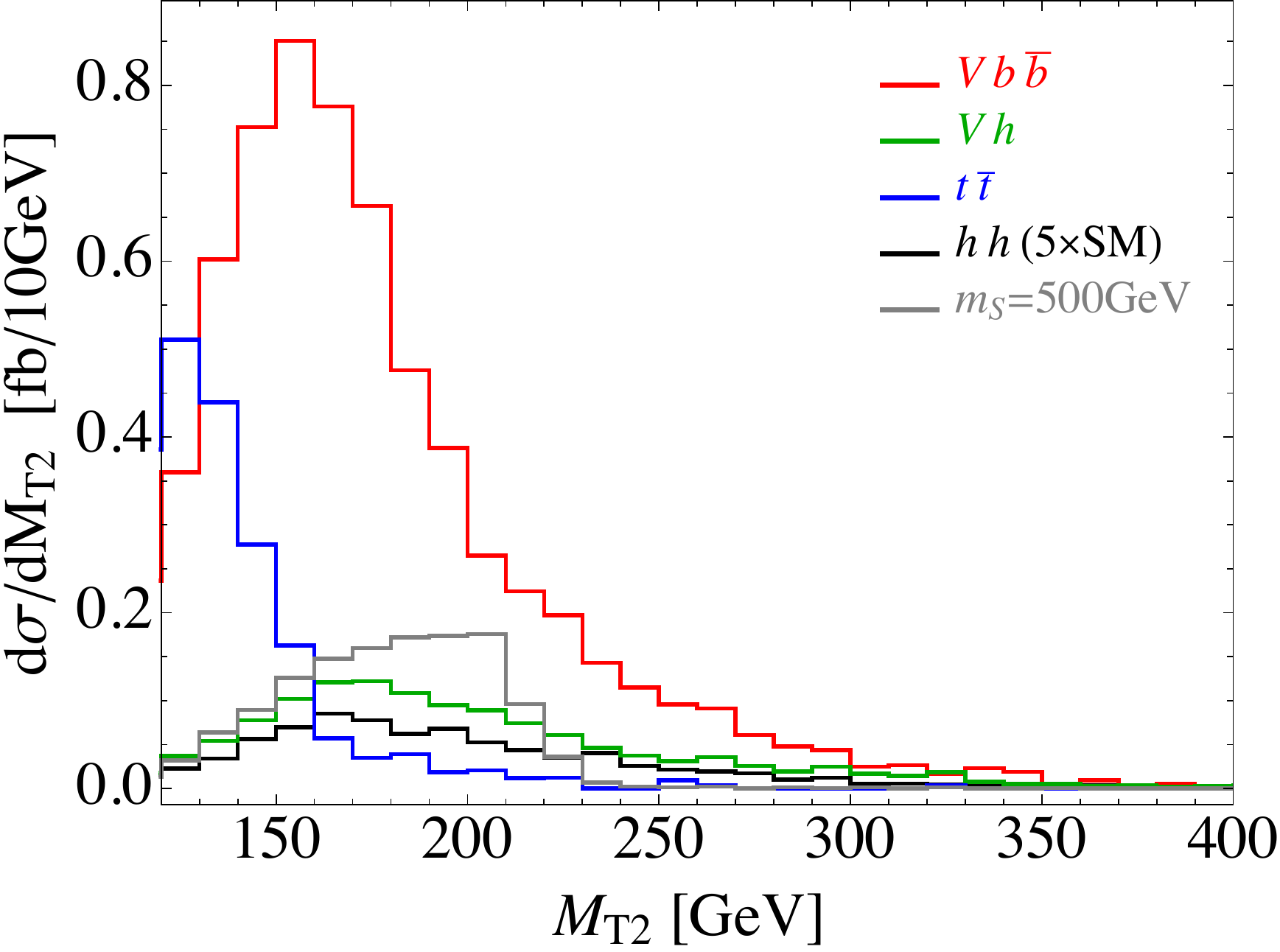}
\caption{
Kinematic distributions for the variables $p_T(b \bar b)$ and $M_{T2}$ before the final event selection. Here we have fixed ${\rm Br}_{\rm inv}=0.2$.
}
\label{fig:kinematics2}
\end{center}
\end{figure*}

In Figure~{\ref{fig:kinematics2}} we show the distributions of transverse momentum of the $b\bar b$  and the stransverse mass variable $M_{T2}$~\cite{Lester:1999tx,Lester:2014yga}~\footnote{We thank Chrisopher G. Lester for clarifications regarding the $M_{T2}$ variable.} following the first set of selections described above.  
To enhance the signal-to-background we exploit the boost of the di-Higgs system by applying two final selections to these variables, demanding $p_{T, b\bar b} > 180$ GeV and $M_{T2}>160$ GeV. 
A cut-flow table with efficiencies after each selection for signal and background is given in Table~\ref{tab:cutflow}.

\begin{table*}[t]
\scriptsize
\begin{tabular}{|c || c | c | c | c | c | c | c |  }
\hline
\hline
	\multirow{3}{*}{}            & \multirow{3}{*}{Signal}   & $W b \bar b$ (no $h$)   & $Zb\bar b$ (no $h$)       & $Wh$                 & $Zh$ \!(1)           & $Zh$ \!(2)                 & $t\bar t$    \\
                                     &                           & $(2b \ell \nu)$         & $(2 b 2\nu/ 2b 2\ell)$    & $(2b \ell \nu)$      & $((2\nu/2\ell)(2b))$ & $((2b)({\slashed{E}}_T))$  & (lep+semi-lep)    \\
\hline
\hline
 $K$-factor                          & \cite{deFlorian:2016uhr}  &  \cite{Cordero:2009kv}  & \cite{Cordero:2009kv}     & \cite{twiki1}        & \cite{twiki1}        & \cite{twiki1}              & \cite{Muselli:2015kba}   \\	
\hline
~$\slashed{E}_T$ trigger + 2$b$+0,1$j$~                        & $1.35\times 10^{-1}$      & $2.81\times 10^{-2}$    & $5.63\times 10^{-2}$      & $1.72\times 10^{-2}$ & $5.21\times 10^{-2}$ & $8.60\times 10^{-2}$       & $7.92\times 10^{-3}$ \\
 \hline
 ~$p_T(b)$~                          & $1.31\times 10^{-1}$      & $2.64\times 10^{-2}$    & $5.12\times 10^{-2}$      & $1.65\times 10^{-2}$ & $4.99\times 10^{-2}$ & $8.10\times 10^{-2}$       & $7.37\times 10^{-3}$ \\	
 \hline
 $m_{bb}$                            & $4.84\times 10^{-2}$      & $7.54\times 10^{-3}$    & $1.50\times 10^{-2}$      & $7.16\times 10^{-3}$ & $2.01\times 10^{-2}$ & $1.73\times 10^{-3}$       & $2.31\times 10^{-3}$ \\
 \hline
 ~$\Delta R(b_1,b_2)$~               & $4.38\times 10^{-2}$      & $5.29\times 10^{-3}$    & $9.95\times 10^{-3}$      & $5.97\times 10^{-3}$ & $1.67\times 10^{-2}$ & $1.32\times 10^{-3}$       & $1.41\times 10^{-3}$ \\
 \hline
 ~$\Delta \phi(bb,{\not \!\! E}_T)$~ & $3.82\times 10^{-2}$      & $5.14\times 10^{-3}$    & $9.56\times 10^{-3}$      & $5.78\times 10^{-3}$ & $1.58\times 10^{-2}$ & $1.24\times 10^{-3}$       & $1.07\times 10^{-3}$ \\
 \hline
  $\slashed{E}_T$                              & $2.35\times 10^{-2}$      & $9.79\times 10^{-4}$    & $2.29\times 10^{-3}$      & $1.62\times 10^{-3}$ & $7.18\times 10^{-3}$ & $6.51\times 10^{-4}$       & $9.50\times 10^{-5}$ \\
 \hline
 $p_T(bb)$, $M_{T2}$                 & $1.44\times 10^{-2}$      & $4.87\times 10^{-4}$    & $8.82\times 10^{-4}$      & $1.21\times 10^{-3}$ & $4.54\times 10^{-3}$ & $3.95\times 10^{-4}$       & $5.73\times 10^{-6}$ \\
 \hline
 \hline
 Scaling                             & $\mu_{hh}$ \!\!\! Br$_{\rm inv}$ \!\!\! (1-Br$_{\rm inv}$)  & 1 & 1 & (1-Br$_{\rm inv}$) & (1-Br$_{\rm inv}$) & Br$_{\rm inv}$ & 1  \\	
 \hline
\end{tabular}
\caption{Cut-flow table for the $b\bar b+{\not \!\! E}_T$ search described in Section~\ref{sec:SM}. Listed in each cell are the efficiencies after the associated cut. The final row displays the scaling of each channel with ${\rm Br}_{\rm inv}$. }
\label{tab:cutflow}
\end{table*}

\begin{figure}\centerline{
\includegraphics[width=0.5\columnwidth]{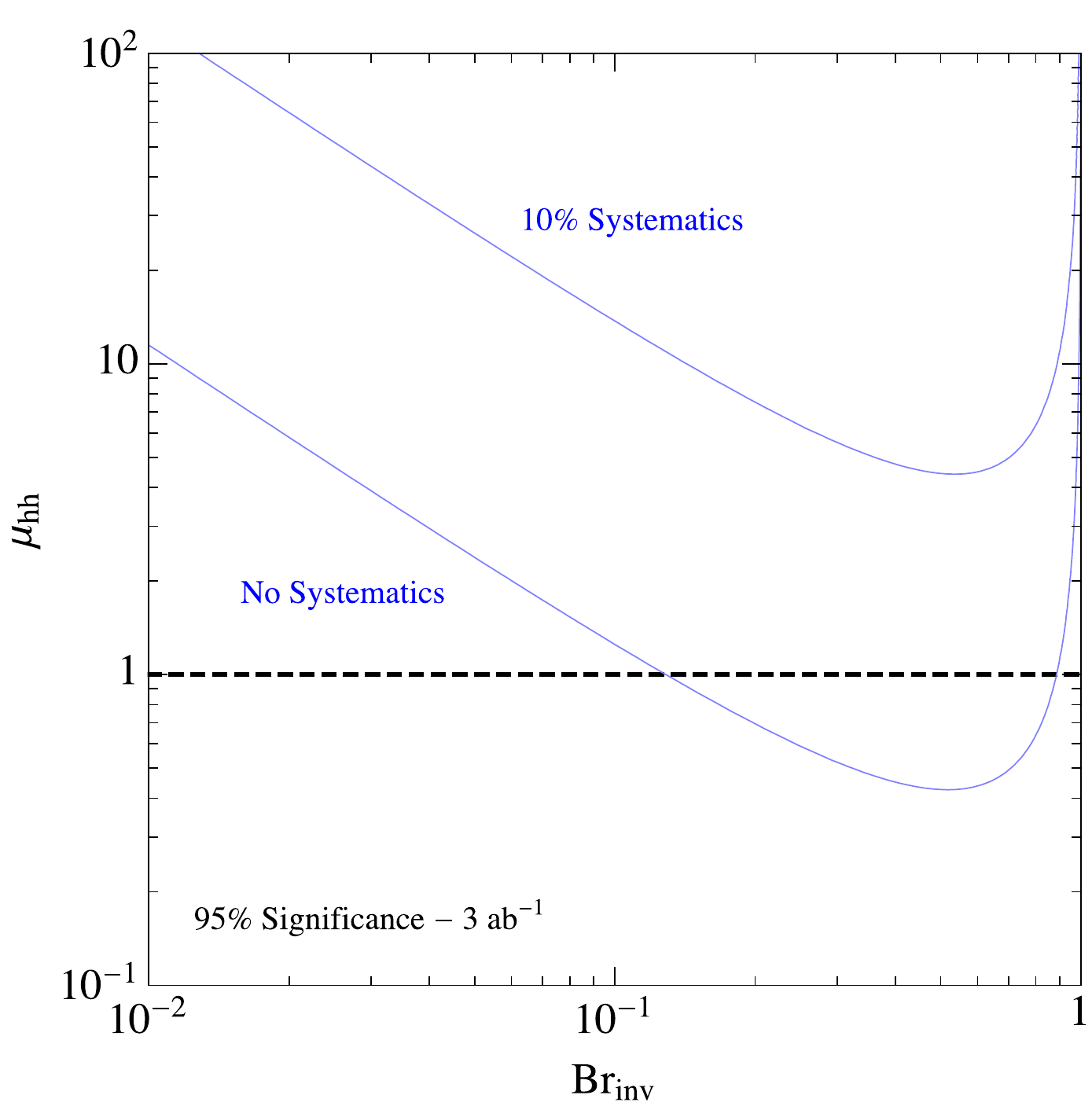}
}
\caption{Reach of the $b\bar b+{\not \!\! E}_T$ search to di-Higgs production at LHC $\sqrt{s} = 14$ TeV with 3 ab$^{-1}$ integrated luminosity. Here we display the $95\%$ significance in the ${\rm Br}_{\rm inv}$ - $\mu_{hh}$ plane for two assumptions on background systematics: 1) statistics dominated, $\gamma_B = \gamma_S = 0$, 
and 2) $10\%$ systematic uncertainty on both signal and background, $\gamma_B = \gamma_S = 0.1$. }
\label{fig:mu-br}
\end{figure}

For each value of invisible branching ratio ${\rm Br}_{\rm inv}$ and di-Higgs signal strength $\mu_{hh}$ we compute the significance for the high luminosity LHC dataset of 3 ab$^{-1}$. We use the following definition for the significance:
\begin{equation}
{\rm Significance} =  \frac{S}{\sqrt{B+ \gamma_B^2 B^2+ \gamma_S^2 S^2}},
\end{equation}
where $S$ ($B$) is the number of signal (background) events after all cuts, and $\gamma_{S,B}$ are the systematic uncertainties in the corresponding rate estimates. In Figure \ref{fig:mu-br} we display the $95\%$ significance in the ${\rm Br}_{\rm inv}-\mu_{hh}$  plane for  two assumptions on systematics: 1) statistics dominated, $\gamma_B = \gamma_S = 0$, 
and 2) $10\%$ systematic uncertainty on both signal and background, $\gamma_B = \gamma_S = 0.1$.  In the statistics dominated case we observe that the LHC can potentially exclude Br$_{\textrm{inv}} \sim 10\%$ at 95\% C.L. for an SM di-Higgs production. On the other hand, if background systematics play an important role, it may be that the LHC will only be able to test this channel if the di-Higgs production rate is enhanced by a factor of order 10 or more, depending on the invisible branching ratio. As an example, considering the case of ${\rm Br}_{\rm inv}=0.2$ and SM production $\mu_{hh}=1$, we find that for $\mathcal{L}=3$ ab$^{-1}$, $S = 298$, $B = 11,231$, $S/B = 0.026$, and $S/\sqrt{B} = 2.82$.

In order to confirm that our cut-based analysis is well optimised we perform a multivariate analysis (MVA) by employing the boosted decision tree (BDT) algorithm and cross-checking the results with a Fisher algorithm. For this purpose we use the TMVA~\cite{2007physics...3039H} framework. We choose 13 kinematic variables with the maximum discriminating power, \textit{viz.} $M_{b_1 b_2}$, $\Delta R(b_1,b_2)$, $p_T^{b_1}$, $p_T^{b_2}$, $\eta^{b_1}$, $\eta^{b_2}$, $\phi^{b_1}$, $\phi^{b_2}$, $\Delta \phi(\slashed{E}_T,b_1b_2)$, $p_T^{b_{1}b_{2}}$, $M_{T2}$, $M_T$, $\slashed{E}_T$, where the indices $1,2$ refer to $p_T$ ordered $b$-jets. While performing the MVA, we carefully treat the issue of overtraining of the signal and background. TMVA performs the Kolmogorov-Smirnov (KS) test to check for any overtraining of the samples. The KS probability for each sample must lie between 0.1 to 0.9. However, in almost all cases, a critical KS probability larger than 0.01~\cite{KS} ensures that the samples are not overtrained. In our analysis, we found that the BDT analysis improves the $S/B$ marginally to $\sim 0.030$. After employing an optimal cut on the BDT variable, we are left with 593 signal and 19466 background events for the same integrated luminosity. This yields a slightly larger $S/\sqrt{B} = 4.19$. Hence we see that the cut-based and the multivariate analyses agree very well and, unfortunately, we are left with a small $S/B$ making it challenging to disentangle the signal particularly when accounting for realistic systematic background uncertainties.

Thus, the limiting factor for this analysis seems to be the strong dependence on the systematic accuracy of the background prediction. It is conceivable that until the end of the runtime of the LHC there will be advances in the reduction of theoretical and systematic uncertainties, possibly after incorporating some data-driven techniques. This might result in an understanding of the background distributions at the level $\mathcal{O}(5)$\%, while the addition of multi-variate reconstruction techniques might improve $S/B$ further. However, realistically, setting a limit on ${\rm Br}_{\rm inv}$ for Standard Model di-Higgs production will remain challenging.

\section{Enhanced Resonant Di-Higgs Production }
\label{sec:resonance}

As seen in the previous section, the LHC has the potential in the long term high luminosity run to probe invisible Higgs decays in the di-Higgs channel, although the case in which the Higgs pairs are produced with the SM rate will likely be challenging. 
However, there are a number of motivated extensions of the SM in which di-Higgs production rate can be significantly enhanced either due to new light colored particles running in the loop, higher dimension operators, or a new heavy di-Higgs resonance. Each of the three mechanisms to enhance the production cross section will result in different kinematics for the $hh$ system \cite{Dolan:2012ac}.

Here we will investigate the prospects of the LHC to detect a new heavy di-Higgs resonance, denoted by $S$, in the case that one of the Higgs particles decays invisibly. Compared to other possible di-Higgs enhancement scenarios, the case of the resonance is perhaps the most interesting from an experimental perspective. 
Not only can the $hh$ production cross section be appreciably increased, but the kinematic properties of the final state particles can be dramatically altered. A heavy resonance with mass $m_S \gg 2 m_h$ will yield highly boosted Higgs bosons, and in the case of the $b \bar b +{\not \!\! E}_T$ channel, highly boosted bottom jets and large missing transverse energy, allowing for a straightforward separation of the di-Higgs signal from the background. 

Our goal in this section will be to demonstrate this fact in a rather model independent fashion. 
The only model-independent constraints come from existing LHC searches for di-Higgs resonances. 
A number of searches have been performed with the 8 TeV data set, including in the 
$4b$, $2b2\gamma$ and $2b 2\tau$ channels~\cite{Aad:2015xja,Khachatryan:2016sey,Khachatryan:2015yea,CMS:2016zxv}. 
Scaling these 8 TeV limits using the ratio of gluon luminosities at 14 TeV and 8 TeV, 
we find the constraints on the 14 TeV resonance cross section, $\sigma(pp\rightarrow S\rightarrow hh)_{14 \, {\rm TeV}}$, range from 
 $25\, {\rm  pb} - 200 \, {\rm fb} $ for resonance masses in the range 200 GeV - 1 TeV. 
More recently, the preliminary results of a search using 13.3 fb$^{-1}$ of $\sqrt{s} = 13$ TeV in the $hh\rightarrow 4 b$ channel have been presented~\cite{ATLAS:2016ixk}. 
This search places a stronger limit on a di-Higgs resonance than the 8 TeV searches, although we note that the analysis is tailored to the case of a spin 2 resonance. Taking this preliminary result at face value, in terms of the 14 TeV resonant di-Higgs cross section, $\sigma(pp\rightarrow S\rightarrow hh)_{14 \, {\rm TeV}}$, this search places a stronger limit than the 8 TeV analyses, ranging from $1\, {\rm  pb} - 50 \, {\rm fb} $ for resonance masses in the range 200 GeV - 1 TeV~\cite{ATLAS:2016ixk}. As an example, for a 500 GeV resonance, the constraint on the di-Higgs resonant cross section is of order 450 fb, or about 12 times larger than the SM di-Higgs production cross section at 14 TeV. We note that these limits assume SM Higgs branching ratios, and thus in our scenario the upper limit on the cross section will be larger by a factor $(1-{\rm Br}_{\rm inv})^{-2}$. 
Finally, we note that existing mono-Higgs searches~\cite{ATLAS:2016thr,CMS:2016xok,mono-higgs1,mono-higgs2} aimed at probing dark matter models do not provide competitive constraints with the the direct di-Higgs searches discussed here.
 
In light of these direct constraints on a di-Higgs resonance, we will here examine the following benchmark scenario
\begin{equation}
m_S = 500\, {\rm GeV}, ~~~~~ \sigma(pp\rightarrow S\rightarrow hh)_{14 \, {\rm TeV}} = 5 \times \sigma^{hh}_{\rm SM}, ~~~~~ \Gamma_S/m_S = 0.01.
\label{eq:benchmark}
\end{equation}
While it is clear that this scenario survives the direct searches described above, it is also straightforward to obtain this enhanced cross section in realistic models containing a di-Higgs resonance while respecting other model-dependent constraints. As one recent example, Ref.~\cite{Dawson:2015haa} has investigated the scalar singlet extension of the SM, finding that a 500 GeV di-Higgs resonance can lead to cross sections with an enhancement of a factor of 5-10 for phenomenologically viable parameter choices. Furthermore, the benchmark scenario (\ref{eq:benchmark}) resembles the ones proposed in \cite{Kotwal:2016tex}, which allow for a strong first-order phase transition within the xSM.
We refer the reader to the literature for other examples of specific models~\cite{Plehn:1996wb,Djouadi:1999rca,Belyaev:1999mx,BarrientosBendezu:2001di,Pierce:2006dh,Arhrib:2009hc,Asakawa:2010xj,Kribs:2012kz,Dawson:2012mk,Dolan:2012ac,Goertz:2013kp,Cao:2013si,Han:2013sga,Nishiwaki:2013cma,Haba:2013xla,Enkhbat:2013oba,Kumar:2014bca,Chen:2014xra,Chen:2014xwa,Chen:2014ask,Bhattacherjee:2014bca,Cao:2014kya,Azatov:2015oxa,Grober:2015cwa,Wu:2015nba,Enkhbat:2015bca,Lu:2015jza,Dawson:2015oha,Etesami:2015caa,He:2015spf,Carvalho:2015ttv,Batell:2015zla,Lu:2015qqa,Batell:2015koa,Dawson:2015haa,Cao:2015oaa,Huang:2015tdv,Agostini:2016vze,Costa:2015llh,Grober:2016wmf,Kang:2016wqi}.
 
We now turn to our analysis of the di-Higgs resonance in the $b \bar b + {\not \! \! E}_T$ channel, focusing on the benchmark in Eq.~(\ref{eq:benchmark}). 
Our analysis chain and search strategy is similar to the one employed in Section~\ref{sec:SM}, with a few minor modifications. For the di-Higgs resonance signal events, we use \texttt{MadGraph5\_aMC@NLO}  with the heavy scalar $hh$ model file, which can be found in Ref.~\cite{MG5resonance}. As a cross check, we also use the Herwig 7 implementation of a di-Higgs resonance~\cite{Bellm:2013hwb}\footnote{We thank Andreas Papaefstathiou for a private version of the code which includes the resonant case.}. 
For the $b \bar b + {\not \! \! E}_T$  search we follow same set of cuts as described in Section~\ref{sec:SM} and listed in Table~\ref{tab:cutflow}, with only two modifications. Namely, we tighten the final selections on the the transverse momentum of the $b\bar b$ system and the stransverse mass to take advantage of the boost of the Higgses coming from the resonance. 

The results of this search are given in Table~\ref{tab:resonance}, where we present the sensitivity to the invisible branching ratio for different assumptions on integrated luminosity and background systematic uncertainties. While the statistics limited case $\gamma_S = \gamma_B = 0$ is likely unrealistic, it gives a sense of the ideal sensitivity of the LHC in this channel. As an example, taking a invisible branching ratio ${\rm Br}_{\rm inv} = 0.1$, we find that for an integrated luminosity of 3 ab$^{-1}$, $S = 1139$, $B = 9084$, $S/B \approx 0.13$, and $S/\sqrt{B} \approx12$. Here we have demanded the final selection cuts to be $p_T^{(bb)} > 180$ GeV and $M_{T2} > 180$ GeV~\footnote{An additional similar hard cut on $\slashed{E}_T$ is also expected to yield similar results. We have not optimized the cut-based analysis for the resonant scenario and have left the optimization on the MVA analysis.}. Provided the background systematics can be controlled, the LHC will be able to discover such a di-Higgs resonance for phenomenologically viable invisible branching ratios. We must also mention that the cut-based analysis for the resonance search can be further optimized. Using a BDT technique, for the same integrated luminosity, we obtain 2802 signal and 13338 backgrounds events after the cut on the BDT variable. This yields $S/B \approx 0.21$ and $S/\sqrt{B} \approx 24$. In order to probe a Higgs invisible branching ratio of 5\% at 90\% CL with this channel, we require an integrated luminosity of 54 fb$^{-1}$ (120 fb$^{-1}$) for zero (5\%) systematic uncertainty. Using the same BDT technique, we find that for the zero (5\%) systematic uncertainty scenario, we are left with $\sim27$ ($\sim 58$) signal and $\sim237$ ($\sim 513$) background events, to probe Br$_{\textrm{inv}}=5\%$ at 90\% CL.

While we have explored the $b\bar b +{\not \!\! E}_T$ channel in this work, it will also be important to consider other channels in which one of the Higgs decays invisibly and the other decays to standard channels. This is particularly true for the case of enhanced production, as can occur with a di-Higgs resonance, since then one can more easily exploit clean channels such as $\gamma\gamma+{\not \!\! E}_T$. Although they have been examined in the context of mono-Higgs dark matter searches~\cite{Carpenter:2013xra,Berlin:2014cfa,ATLAS:2016thr,CMS:2016xok,mono-higgs1,mono-higgs2}, it will be important to revisit these channels with the aim of optimizing the search strategies to Higgs pair production. Along these lines, it would be interesting to examine the potential of a future 100 TeV hadron collider, where production rates in these channels could be larger by a factor of 50 or so. We leave this work to future study.

\bgroup
\renewcommand{\arraystretch}{1.2}%
\begin{table}
  \begin{tabular}{| c c   c | c  c |}
  \hline
& \multicolumn{2}{c|}{} & \multicolumn{2}{c|}{\footnotesize Integrated Luminosity} \\ [-10pt]
& \multicolumn{2}{c|}{} & ~~~300 fb$^{-1}$~~~  & ~~~3 ab$^{-1}$~~~   \\  [0pt]
\hline
\hline
&  \multirow{2}{*}{\begin{sideways} {\footnotesize Systematics~~ }\end{sideways}} &  ~~0$\%$~~  & 4$\%$ &  1$\%$  \\ [10pt]
\cline{4-5}
& & ~~10$\%$~~  &  18$\%$ &  17$\%$  \\ [10pt]
\cline{4-5}
\hline
  \end{tabular}
  \caption{Sensitivity to invisible branching ratio for the di-Higgs resonance benchmark in Eq.~(\ref{eq:benchmark}) at 95$\%$ significance for 300 fb$^{-1}$ and 3 ab$^{-1}$ at 14 TeV LHC.}
  \label{tab:resonance}
\end{table}
\egroup

\section{Other Exotic Higgs Decays in Higgs Pair Production}
\label{sec:exotic}

Having discussed in detail the case of invisible Higgs decays, we now turn to discuss the potential of a few more theoretically motivated exotic decay modes in di-Higgs production. Our aim in this section is not to be rigorous but only to lay an outline for certain searches which can serve as a set of guidelines for future studies. We will mostly follow the results given in Ref.~\cite{Curtin:2013fra}, which performed a methodical study of the exotic decays in single Higgs production when quoting existing limits or future sensitivities for particular exotic decay branching ratios. As with the invisible channel, it is most likely that any exotic decays would first be detected in single Higgs production channels due to the larger rate, except perhaps in cases when the backgrounds are easier to handle for the di-Higgs case or in the instance that di-Higgs production is enhanced. 
We will focus on the scenarios in which the Higgs decays to a pair of light (pseudo)scalars which in turn decay to fermions or to gluons/photons. 
Some models where such decays are allowed are a singlet extension of the two Higgs doublet models (2HDM+S)~\cite{Peccei:1977hh,Haber:1984rc,Kim:1986ax}, extensions of SM with hidden light gauge bosons~\cite{Gopalakrishna:2008dv}, the R-symmetry limit of the Next to Minimal Supersymmetric Standard Model (NMSSM)~\cite{Cao:2013gba} and the Little Higgs model~\cite{Surujon:2010ed} to name a few. 
We categorize the proposed searches under two broad categories, \textit{viz.}, exotic Higgs decays with a relatively large branching ratio and ones with a clean signal.

One interesting possibility is the Higgs decaying to 4$b$ like $h \to X X \to 4 b$~\cite{Ellwanger:2003jt,Ellwanger:2005uu,Cao:2013gba}. In all of the aforementioned models one can find parameter regions where Br($X \to b \bar{b}$) can be large. In Ref.~\cite{Curtin:2013fra}, the authors predict the 2$\sigma$ sensitivity of Br($h \to XX \to 4b$) $=$ 0.1 (0.2) with 300 (100) fb$^{-1}$ integrated luminosity in the kinematically allowed region of $X$. In this context if we consider a di-Higgs production with one of the Higgs decaying in this exotic mode and another decaying invisibly, we will expect around $\mathcal{O}(1000)$ events for Br($h \to \slashed{E}_T$) $\sim$ 0.2, before applying any selection cuts. Here we have assumed that the sensitivity of Br($h \to XX \to 4b$) does not change appreciably on going from 300 fb$^{-1}$ to 3 ab$^{-1}$. However, while a $6b$ final state can have a large rate, it will be a challenging task to reconstruct the Higgs given the many combinatorial possibilities. Another channel which can prove promising is $4b+2\ell+\slashed{E}_T$, where $\ell=e,\mu$ , with the other Higgs decaying to $2\ell + \slashed{E}_T$ via $WW^*, ZZ^*$ or $\tau^+ \tau^-$. At 3 ab $^{-1}$, one can expect $\mathcal{O}(100)$ events prior to selection cuts. Clean channels, such as $4b+2\gamma$ are rate limited at LHC with SM production, but they could be interesting in the case of enhanced di-Higgs production or also at a future 100 TeV hadron collider. 
All numbers quoted thus far take into account tagged $b$-jets with a tagging efficiency of 70\%. One must note here that the limit on Br($h \to XX \to 4b$) has been obtained by studying the $Wh$ production mode for the single Higgs because of lesser background. We have used the same limit to estimate the number of events for a di-Higgs produced via gluon fusion. We must also mention that $m_X \lesssim 30$ GeV produces merged $b$-jets and for such cases jet substructure algorithms can be extremely useful.

Another channel which can provide an interesting signature involves an exotic decay of the Higgs as $H \to a a \to 2 b 2 \tau$, where $a$ is a light (pseudo-)scalar the preferentially decays to third generation fermions. In certain models such as the NMSSM or Little Higgs, when $2 m_b < m_a < m_h/2$, the Higgs can have relatively large branching ratio to $aa$, and furthermore the couplings of $a$ to SM fermions can be roughly proportional to the SM Yukawa couplings. 
There is no existing analysis that places a strong limit this branching ratio. 
However, Ref.~\cite{Curtin:2013fra} predicts sensitivity to Br($h \to a a \to 2 b 2 \tau$) $\simeq$ 0.15 at 100 fb$^{-1}$.  
One can envision a search in the di-Higgs channel in which one of the Higgs decays to $2b2\tau$ while the other decays to $2b$, leading to $4b2\tau$ final state. Reconstructing the individual $h$ and $a$ resonances may help overcome the combinatoric challenge posed by four bottom quarks. The rates in these channels can be similar to those described above for the exotic decay $h\rightarrow 4b$.

We know that in the SM the channel Higgs decaying to 4 jets is only possible via $WW^*/ZZ^*$. Out of the 4 jets only one pair comes from an on-shell particle. However in an extended scenario like the one described above we can have a process like $h \to a a \to 4 j$ and where we can reconstruct both pairs of jets. For $m_a \lesssim 5$ GeV one can constrain Br($h \to a a \to 4j$) $=$ 10\% at 300 fb$^{-1}$. One can study similar di-Higgs channels to the ones mentioned above for the exotic decay $h \to 4b$. 
Here we will not have to pay the price of $b$-tagging. However, we will have need to reconstruct both the $a$ pseudoscalar resonances. The other Higgs can decay to a pair of bottoms quarks, $2 \ell + \slashed{E}_T$, or perhaps even two photons and still give a sizable event yield depending on the di-Higgs production mechanisms. A similar exotic channel can be $h \to 2 \gamma + 2j$.  The current upper limit on Br($h \to a a \to 2 \gamma + 2 j$) $=$ 0.04, suggesting the potential for a few thousand events in the di-Higgs channel $2b2j2\gamma$ at the high-luminosity LHC.

It has also been shown that in the Peccei-Quinn limit of NMSSM one can expect exotic decays like $h \to \chi_1 \chi_2 \to 2 b + \slashed{E}_T$, where the $b\bar{b}$ may or may not be resonant. The limit obtained in this analysis is Br($h \to \chi_1 \chi_2 \to 2 b + \slashed{E}_T$) $=$ 0.2 at 300 fb$^{-1}$. Assuming the other decay mode is $2 l + 2 \nu$, one can expect on the order of a few hundred
events for a high-luminosity LHC. For this analysis, the $M_{T2}$ variable has the potential to be very useful. The other Higgs can also decay to a $b\bar{b}$ pair and provide an even bigger rate. Along similar lines $h \to \chi_1 \chi_2 \to 2 \mu + \slashed{E}_T$ can prove very useful. The present bound on this branching ratio from 8 TeV is 0.07. A final state like $2 \mu + \slashed{E}_T + 2 b$ can have a good potential. Lastly, we would like to discuss another possible final state where one of the Higgses decays to $2\gamma + \slashed{E}_T$ and the diphoton pair can either be resonant or non-resonant. From 8 TeV searches, the authors of Ref.~\cite{Curtin:2013fra} have placed a limit on the Br($h \to 2 \gamma + \slashed{E}_T$) at 4\%. Assuming that the 14 TeV searches can constrain it to a percent level, one can still expect $\mathcal{O}(1000)$ events prior to selection cuts at 3 ab$^{-1}$ with the other Higgs decaying to $b\bar{b}$.

In this section we have tried to motivate both theorists and experimentalists to examine possible exotic modes in the di-Higgs channel. We have surveyed only a handful of channels, but there are other possibilities, including for example, the case of displaced exotic Higgs decays, or events in which both of the Higgs particles decays through an exotic mode.
Our considerations here are only very cursory, and indeed it remains to be seen whether any of the particular channels can be observed after performing a realistic collider simulation that accounts for backgrounds and detector efficiencies. While it may prove to be challenging, the physics case for exotic decays is well motivated as is the search for Higgs pair production, and it is thus important to understand the extent to which the LHC can probe such exotic channels. 

\section{Conclusion}
\label{sec:conclusion}

The search for Higgs pair production will continue to be an important enterprise for the LHC experiments moving forward due to its connection to the structure of the scalar potential. It is important to understand how BSM physics can affect di-Higgs detection. In this paper we have investigated the consequences of invisible Higgs decays on the di-Higgs channel. Invisible Higgs decays are a generic feature in many extensions of the Standard Model and are now being actively searched for at the LHC in processes involving a single Higgs particle. The current constraints on the invisible branching ratio are still quite weak, with ${\rm Br}_{\rm inv} \lesssim 0.25$ still allowed by the LHC combinations depending on the assumptions regarding the modifications to Higgs branching ratios. While it is clear that invisible Higgs decays would most likely be first seen in single Higgs production modes, it is nevertheless worth asking if they can also eventually be observed in Higgs pair production. 

With this motivation, we have described a search for Higgs pair production in the $b\bar b+{\not \!\! E}_T$ channel. In the case of SM di-Higgs production, while our search potentially has statistical sensitivity, we find that it will be challenging to probe invisible decays in this channel for realistic estimates of the background systematic uncertainties and phenomenologically allowed invisible branching ratios. We have also considered the interesting case of a di-Higgs resonance. Such a resonance can significantly enhance the production rate and lead to more distinctive signal kinematics. Depending on the invisible branching ratio, as well as the mass and the cross section of the new state, such a di-Higgs resonance may be discoverable in LHC Run-II with our proposed search. 

This is a first investigation into the consequences of exotic decays in Higgs pair production. While the invisible channel is perhaps the simplest and certainly one of the best motivated, there are clearly many other possibilities worth considering. While it seems likely that that such exotic Higgs decays would first be observed in single Higgs production channels, a basic question of interest is whether or not such exotic decays could actually enable earlier discovery of Higgs pair production than the standard channels. We have given a brief survey of possibilities here with the hope of encouraging both experimentalists and phenomenologists to  examine their physics potential more carefully and to ultimate perform searches in promising channels. We look forward to continued activity along this direction. 

\section*{Acknowledgements}
We thank Tae Min Hong, Chrisopher G. Lester, Tanumoy Mandal, Alexandre Mertens, Michele Selvaggi and Emanuele Re for helpful discussions. We are especially grateful to Andreas Papaefstathiou for help regarding implementation of di-Higgs in Herwig 7. SB would like to extend his thanks to Shilpi Jain for helpful discussions and technical help during various stages of this project.
BB is supported in part by the U.S. Department of Energy under grant No. DE-SC0015634. The work of BB was performed in part at the Aspen Center for Physics, which is supported by National Science Foundation grant PHY-1066293. 
SB acknowledges the support of the Indo French LIA THEP (Theoretical high Energy Physics) of the CNRS. MS is supported in part by the European Commission through the “HiggsTools” Initial Training Network PITN-GA-2012-316704.

\bibliography{refs}

\end{document}